\documentstyle[epsfig,amssymb,aps,prb]{revtex}

\begin{document}
\title{Excitons and charged excitons in semiconductor quantum wells}
\author{C. Riva\cite{clara} and F. M. Peeters\cite{francois}}
\address{Departement\ Natuurkunde, Universiteit Antwerpen (UIA),
B-2610 Antwerpen.}
\author{K. Varga\cite{kvarga}}
\address{Department of Physics, Argonne National Laboratory,
9700 South Cass Avenue, Argonne, 60439 Illinois, USA.}
\date{\today }
\maketitle

\begin{abstract}
A variational calculation of the ground state energy of neutral excitons and
of positively and negatively charged excitons (trions) confined in a single
quantum well is presented. We study the dependence of the correlation energy
and of the binding energy on the well width and on the hole mass. The
conditional probability distribution for positively and negatively charged
excitons is obtained, providing information on the correlation and the
charge distribution in the system. A comparison is made with available
experimental data on trion binding energies in GaAs, ZnSe and CdTe based
quantum well structures which indicates that trions become localized with
decreasing quantum well width.

\end{abstract}
\nopagebreak
\twocolumn

\section{Introduction}

Negatively (X$^{-}$) and positively (X$^{+}$) charged excitons, also called
{\em trions}, have been the subject of intense studies in the last years,
both experimentally and theoretically. The stability of charged excitons in
bulk semiconductors was proven theoretically by Lampert\cite{Lampert} in the
late fifties, but only recently they have been observed in quantum well
structures: first in CdTe/CdZnTe by Kheng {\it et al.}\cite{Kheng} and
subsequently in GaAs/AlGaAs.\cite{Finkelstein,Shields}

After the initial work by Lampert charged excitons in bulk semiconductors
\cite{Stebe75} as well as in an exactly two-dimensional (2D) configuration
\cite{Stebe89} were systematically studied theoretically. These studies
revealed that, due to the confinement, the 2D charged excitons have binding
energies which are an order of magnitude larger than the charged excitons in
the corresponding bulk materials. Apart from these two early studies several
works were recently published on charged excitons in a high magnetic field,
\cite{Chapman,Whittaker} where one is allowed to use the single particle
Landau Level approximation, or in the presence of an electric field.\cite
{Dujardin}

In order to limit the computational time, the previous theoretical
calculations used approximations and/or simplifications in the Hamiltonian,
e.g. replacing the true Coulomb interaction by an average interaction, or in
the wave function, i.e. neglecting the correlation among the particles in
one or more spatial directions. Because the binding energy of the trions is
very sensitive to the correlation between the different particles, it would
be interesting to have a full calculation in order to evaluate the
approximations that have been made. We present here a calculation of the
ground state energy for the exciton and the charged exciton based on the
stochastic variational approach that fully includes the Coulomb interaction
among the particles (for preliminary results using this method see Ref. %
\onlinecite{ascona}). The use of the stochastic method allows us to handle a
big number of variational parameters in a reasonable time and to
systematically increase the accuracy of our solution.

In the present paper we study the X$^{-}$ and X$^{+}$ systems in a single
quantum well (QW) with a finite height of the potential barrier. In the
first section we present the Hamiltonian of the problem. In the second
section we discuss the dependence of the charged exciton correlation energy
on the well width and on the hole mass. We compare our results with those of
St\'{e}b\'{e} {\it et al.}\cite{Stebe97}, where a variational technique with
a 66-terms Hylleras trial wave function was used. In this section we also
present our results for the binding energy of the X$^{-}$ and X$^{+}$ and we
discuss the pair correlation functions and the probability density of the
system. Our results are compared with available experimental data from the
literature. In the last section we summarize our results and give our
conclusions.

\section{The model}

In this section we present the Hamiltonian describing a charged exciton and
we discuss the technique which we used to solve it. In particular, we focus
on the Hamiltonian of a negatively charged exciton. The positively charged
exciton Hamiltonian can be easily obtained from the X$^{-}$ by replacing the
electrons by holes and the hole by an electron.

The Hamiltonian of a negatively charged exciton in a quantum well is, in the
effective mass approximation, given by
\begin{equation}
\widehat{H}=T_{1e}+T_{2e}+T_{h}+V_{C}+V_{1e}+V_{2e}+V_{h},  \label{HAM}
\end{equation}
where $1e$, $2e$ indicate the electrons and $h$ the hole; $V_{ie},$ $V_{h}$
are the quantum well confinement potentials; $T_{i}$ is the kinetic energy
operator for particle $i$,
\begin{equation}
T_{i}=\frac{\overrightarrow{p}_{i}^{2}}{2m_{i}},
\end{equation}
with $m_{i}$ the mass of the $i$-th particle; $V_{C}$ is the sum of the
Coulomb electron-electron and electron-hole interactions,
\begin{equation}
V_{C}=\frac{e^{2}}{\varepsilon }\left( \frac{1}{|\overrightarrow{r}_{1e}-%
\overrightarrow{r}_{2e}|}-\frac{1}{|\overrightarrow{r}_{1e}-\overrightarrow{r%
}_{h}|}-\frac{1}{|\overrightarrow{r}_{2e}-\overrightarrow{r}_{h}|}\right) ,
\end{equation}
with $e$ the elementary charge and $\varepsilon $ the static dielectric
constant. For a GaAs/Al$_{x}$Ga$_{1-x}$As quantum well the heights of the
square well confinement potentials are $V_{ie}=0.57\times (1.155x+0.37x^{2})$
eV for the electrons and $V_{h}=0.43\times (1.155x+0.37x^{2})$ eV for the
hole.

The Hamiltonian is then solved using the stochastic variational method.\cite
{varga} The trial function, for the variational calculation, is taken as a
linear combination of ``deformed'' correlated Gaussian functions,
\begin{eqnarray}
&&\phi _{0}(\overrightarrow{r}_{1e},\overrightarrow{r}_{2e},\overrightarrow{r%
}_{h})=\sum_{n=1}^{K}C_{n0}\Phi _{n0}(\overrightarrow{r}_{1e},%
\overrightarrow{r}_{2e},\overrightarrow{r}_{h}),  \label{wave-function} \\
&&\Phi _{n0}(\overrightarrow{r}_{1e},\overrightarrow{r}_{2e},\overrightarrow{%
r}_{h})={}{\cal A}\left\{ \exp \left[ -{\frac{1}{2}}\sum_{%
\renewcommand{\arraystretch}{0.6}{\
\begin{array}{c}
\scriptstyle{i,j\in \{1e,2e,h\}} \\
\scriptstyle{k\in \{x,y,z\}}
\end{array}
}}A_{nijk0}r_{ik}r_{jk}\right] \xi(1,2,3)\right\} ,  \nonumber
\end{eqnarray}
where $r_{ik}$ gives the position of the $i$-th particle in the direction $k$%
; ${\cal A}$ is the antisymmetrization operator and $\{C_{n0},A_{nijk0}\}$
are the variational parameters, and $\xi(1,2,3,)$ is the spin function.
The `0' in Eqs. (\ref{wave-function})
refers to the ground state. Note that in contrast with the ``classical''
correlated Gaussians here, the parameter $A_{nijk0}$
which expresses the correlation among the particle $i$ and the particle $j$
in the direction $k$, is allowed to be different from the parameter $%
A_{nijk^{^{\prime }}0}$ which couples the same two particle $i$ and $j$ in a
different direction $k^{^{\prime }}.$ This additional degree of freedom in
the calculation allows us to take into account the asymmetry introduced in
the 3D space by the presence of the quantum well. The dimension of the
basis, $K$, is at first increased until the energy is accurate to the second
digit, a typical value of $K$ in this calculation is 300, and then is
refined to increase the accuracy. The refinement is made by replacing the
n-th state with a new state, i.e. with a state built using new parameters $%
C_{n0},A_{nijk0}$ in such a way that it lowers the total energy. The process
is reiterated multiple times for all the $K$ states, until the energy
reaches the desired accuracy. One can get faster convergence by taking into
account the cylindrical symmetry choosing $A_{nijx0}=A_{nijy0}$.

\section{Theoretical results}

\begin{figure}[tbp]
\begin{center}
\epsfig{file=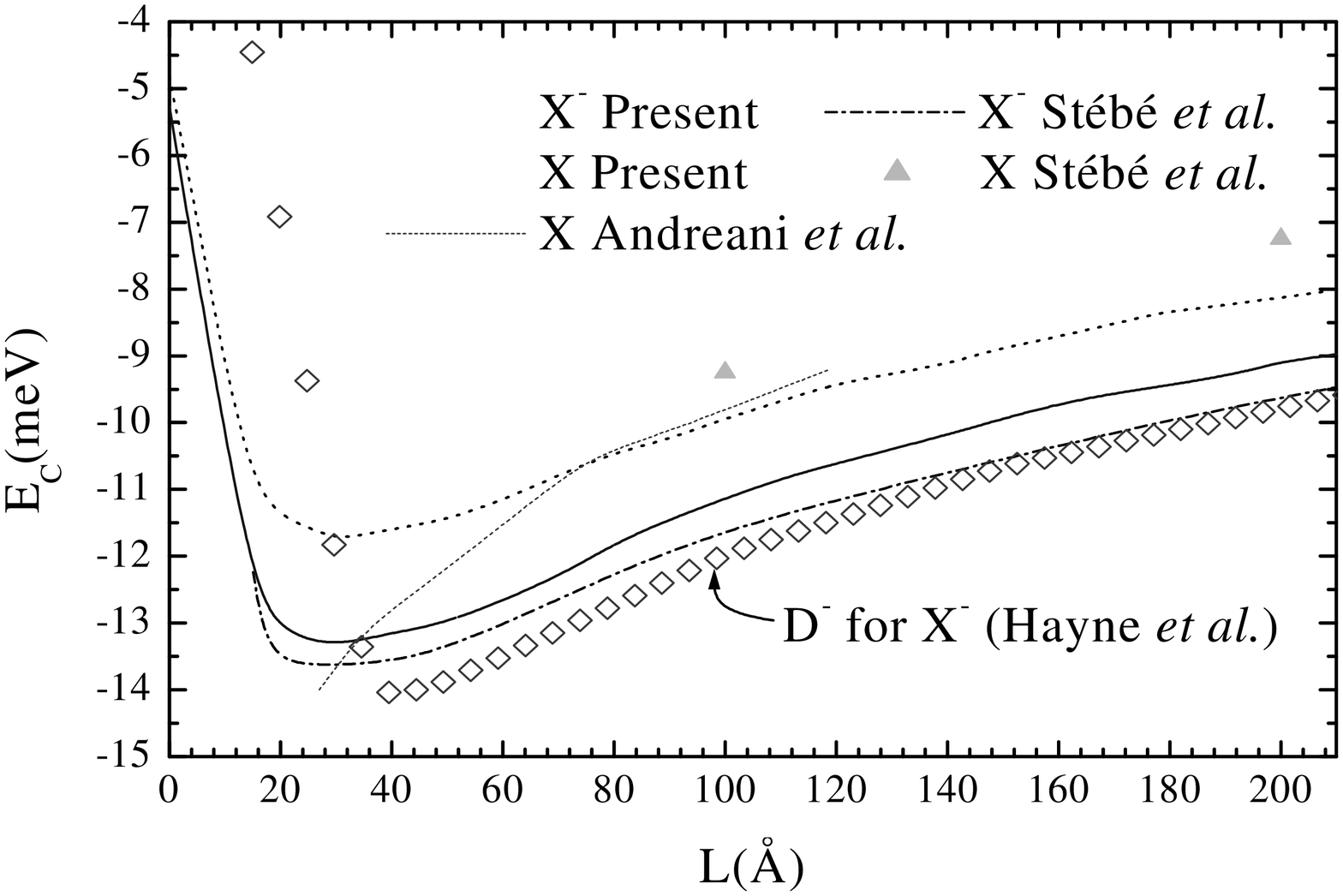,width=8cm}
\end{center}
\caption{The correlation energy of the exciton and the negative charged
exciton vs. the quantum well width. }
\label{Comp-stebe}
\end{figure}

\noindent The correlation energy of a charged exciton is defined as
\begin{eqnarray}
\text{E}_{C}\text{(X}^{-}\text{)} &=&\text{E(X}^{-}\text{)}-\text{2E}_{e}-%
\text{E}_{h}\text{,} \\
\text{E}_{C}\text{(X}^{+}\text{)} &=&\text{E(X}^{+}\text{)}-\text{2E}_{h}-%
\text{E}_{e}\text{,}
\end{eqnarray}
with E(X$^{\pm }$) the energy level of the charged exciton and E$_{e}$ and E$%
_{h}$ the energy levels of the free electron and hole, respectively, in the
quantum well. Thus, E$_{C}$ is the energy due to the Coulomb interaction
between the charged particles. We discuss here the results obtained for a
GaAs/Al$_{x}$Ga$_{1-x}$As quantum well with $x=0.3$, where the value of the
masses used are $m_{e}=0.0667m_{0},$ $m_{hh}=0.34m_{0},$ i.e. GaAs masses,
with $R_{y}=\hbar /m_{e}a_{B}=5.79$ meV, the donor effective Rydberg, and $%
a_{B}=\hbar ^{2}\varepsilon /m_{e}e^{2}=99.7$ \AA , the donor effective Bohr
radius. The results for the correlation energy of the exciton and the X$^{-}$
are shown in Fig. \ref{Comp-stebe} and compared with the theoretical results
of others. We observe that the correlation energy of the exciton, E$_{C}$%
(X)=E(X)$-$2E$_{h}-$E$_{e},$ (dotted line in Fig. \ref{Comp-stebe})
increases in absolute value for well widths up to L$=30$\AA , \ were it
reaches a minimum E$_{C}$(X)$=-11.7$ meV. For L$<30$\AA\ and with decreasing
L the exciton correlation energy decreases in magnitude due to the fact that
the electron, and to a lesser extent, the hole wave functions spill over
into the barrier material of the quantum well. Consequently the exciton
becomes more extended in the $z$-direction and the Coulomb interaction among
the particles is diminished. At L=0 we obtain E$_{C}(X)$=$-$4.80 meV which
compares with the correlation energy of an exciton in bulk GaAs, i.e. E$%
_{C}^{3D}$(X)=$-$4.84 meV. For L%
\mbox{$>$}%
30\AA\ and increasing L the correlation energy decreases in magnitude with
L, which is due to the fact that the electron and the hole are more extended
in the $z$-direction. In the limit L $\rightarrow \infty $ we recover the 3D
exciton in bulk GaAs.

The correlation energy of the negatively charged exciton has the same
qualitative L-dependence as the exciton. It reaches a minimum at about 30
\AA \thinspace\ with E$_{C}$(X$^{-}$)=$-$13.2 meV. For L%
\mbox{$>$}%
30\AA\ it proceeds almost parallel to E$_{C}$(X), in the region shown. For
the X$^{-}$ we obtain E$_{C}^{3D}$(X$^{-}$)=$-$4.95 meV as 3D correlation
energy.

We compare our results for the X in GaAs/Al$_{0.3}$Ga$_{0.7}$As to the ones
reported in {\it \ }Ref.\onlinecite{andreani2} (short-dashed curve in Fig.
\ref{Comp-stebe}), which are derived using the theory in Ref. %
\onlinecite{andreani1}. The theory of Andreani and Pasquarello\cite
{andreani1} also includes the non isotropy of the masses, the
non-parabolicity of the conduction band and the dielectric constant
mismatch. Moreover the values for the heights of the potential barrier used
are slightly different with respect to ours. However, a comparison between
the two calculations can still be made. We observe that our results and the
ones in Ref. \onlinecite{andreani2}, are very close in the range 70-120 \AA
. For L%
\mbox{$<$}%
70\AA , our values for the correlation energy are much smaller, in absolute
value, than the ones reported in Ref.\onlinecite{andreani2}, which is due to
the band non-parabolicity which is known from Ref. \onlinecite{andreani1} to
be the major factor responsible for the steep decrease of the correlation
energy at small quantum well widths. We want now to estimate, although in a
naive way, the effect of band non-parabolicity on our results. From Ref.%
\onlinecite{andreani1} we estimate that for a quantum well L=30\AA , and $x$%
=0.3 the parallel mass of the electron is about 0.08$m_{0}$. If we consider,
in a very simplified picture, that: a) the contribution to the correlation
energy strongly depends on the mass along the growth direction, namely the
confinement energy, which has been subtracted out in the correlation energy,
and that b) the energy of the exciton is largely dominated by the mass of
the electron, through the change in the Rydberg which re-scales the energy.
Such a procedure gives $E_{C}$=$-$13.9 meV, for the case of a quantum well
of width L=30\AA ,\ which compares very well with the value $-$14 meV given
for a quantum well of width L=27 \AA\ in Ref. \onlinecite{andreani2}.

Both for the cases of an X and of an X$^{-}$ in a GaAs/Al$_{0.3}$Ga$_{0.7}$%
As quantum well we compare our results to the one obtained in Ref. %
\onlinecite{Stebe97}. Our calculation gives qualitatively the same
correlation energy both for the X and for the X$^{-}$ as compared to the one
given by St\'{e}b\'{e} {\it et al.}\cite{Stebe97} However while for the
exciton we find that the correlation energy is lower than the one obtained
in Ref. \onlinecite{Stebe97}, thus indicating that the Coulomb correlation
is more fully included in our approach, for the negatively charged exciton
our approach gives a higher correlation energy (about 4$\%$ for L=100\AA ).
The latter can be understood as follows: in Ref. \onlinecite{Stebe97} the
Coulomb potential along the $z$-direction was approximated by an analytical
form and Hylleras-type functions were used for the wave function. They
calculated the Coulomb potential matrix between any two basis states ($s1,s2$%
) by integrating it over the $\rho $-plane thus obtaining a potential matrix
$V_{s1,s2}(z)$. Then $V_{s1,s2}(z)$ was replaced by the analytical
expression $-\gamma /(\delta +|z_{1e}-z_{h}|)-\gamma /(\delta
+|z_{1e}-z_{h}|)+\gamma /(\beta +|z_{1e}-z_{2e}|)$ where $\gamma ,$ $\delta $
and $\beta $ were determined in such a way that it reproduces the correct
behavior of the Coulomb potential matrix in the limit of zero and infinite
distance between the particles in the $z$-direction. This approximation
leads, as the authors of Ref. \onlinecite{Stebe97} noted, to an error in the
exciton correlation energy which was estimated to increase its absolute
value by approximately 5\%. Our present results for the exciton energy are
about 8\% lower than those of St\'{e}b\'{e} {\it et al.}\cite{Stebe97} For
the charged exciton energy the authors of Ref.\onlinecite{Stebe97} did not
report an estimate of the error which was introduced through the
approximations made. However, we find a smaller correlation energy of about
4\% as compared to those of Ref. \onlinecite{Stebe97}. We think that this
result is not in conflict with the one obtained for the exciton. Indeed,
while in the exciton case in Ref. \onlinecite{Stebe97} \ only the attractive
interaction between the electron and hole was underestimated, for the
charged exciton case the repulsive interaction between the electrons will
also be underestimated. The difference is that the former interaction has
the effect of increasing the bonding of the particle while the latter has
the effect to diminish it. Our result indicates that a larger error is made
in the electron-electron repulsive interaction in Ref. \onlinecite{Stebe97}
as compared to the error in the electron-hole attractive interaction.

The approximation by St\'{e}b\'{e} {\it et al.}\cite{Stebe97} consisted in
averaging the wave function in the xy-plane in order to find an effective
Hamiltonian describing the exciton and the trion in the z-direction. This is
similar to an adiabatic approximation which is valid when the motion in the
xy-plane is faster then the one in the z-direction. We believe that it is
more natural to do the reverse and average over the particle motion in the
z-direction which is due to the quantum well confinement and which will be
much faster. Such an approach is equivalent to neglect the particle-particle
correlation along the z-direction which we expect to be valid when $%
E_{e(h)}>>E_{C}^{X,X^{-}}$. For both the exciton and the trion this relation
is satisfied for L$<$150 \AA . Averaging Eq.(\ref{HAM}) over $z$ we obtained
an effective 2D Hamiltonian, in which the effective Coulomb potential was
replaced by the analytical form $e^{2}/\epsilon \lbrack \lambda +(\vec{\rho}%
_{i}-\vec{\rho}_{j})^{2}]^{1/2}$, where $\lambda $ was obtained by fitting
this analytical form to the numerical results for the effective Coulomb
interaction. The correlation energy for the exciton and the charged exciton
is in this case lower than the one we obtain with our more exact calculation
presented above, e.g. in the frame of the model presented in this paper we
find E$_{C}$(X)$=-$10.1meV and E$_{C}$(X$^{-}$)$=-$10.9meV for a 100 \AA\ %
wide quantum well, and E$_{C}$(X)$=-$10.7meV and E$_{C}$(X$^{-}$)$=-$11.6meV
for a 80 \AA\ wide quantum well, while using the screened 2D Coulomb
potential we found E$_{C}$(X)$=-$10.4meV and E$_{C}$(X$^{-}$)$=-$11.4meV for
a 100 \AA\ wide quantum well and E$_{C}$(X)$=-$11meV and E$_{C}$(X$^{-}$)$=-$%
12.2meV for a 80 \AA\ wide quantum well. Consequently, such an approach
leads to larger correlation energies and also to slightly larger binding
energies.

In Fig. \ref{Comp-stebe} we also report the result of a simplified model
(open diamonds) for the study of the energy of a trion which we proposed in
Ref. \onlinecite{Manus}. This model is derived from the one used for a D$%
^{-} $ system,\cite{clara-d} and it assumes that the hole is fixed at the
center of the well, i.e. it has an infinite mass. The effect of the hole is
reflected in the renormalization of the mass of the electron, i.e. m$_{e}$
is replaced by the reduced mass $\mu =m_{e}m_{h}/(m_{e}+m_{h})$. This model
gives for the correlation energy of the charged exciton results that are, at
first, surprisingly close to the one obtained by St\'{e}b\'{e} {\em et al.},
\cite{Stebe97} at least down to well widths of about 40 \AA . This seems to
suggest that the procedure of averaging the potential in the plane adopted
in Ref. \onlinecite{Stebe97} is almost equivalent to localize the hole in
the $\rho $-plane. For smaller well width the magnitude of the correlation
energy becomes much larger as compared both to our present result and to the
one of Ref. \onlinecite{Stebe97}. As shown in Fig. \ref{Comp-stebe} the
result we obtain in the L=0-limit is dramatically different from the one
found in the present work: E$_{C}$=$-$5.2 meV. The reason is that for small
well widths the hole may no longer be considered as a '' fixed'' particle
and the penetration of the hole in the barrier can no longer be neglected.

To prove further the accuracy of our calculation and to check the quality of
our wave function for the X$^{-}$ we calculated the virial which is defined
as
\[
{\em v}\text{=}2\frac{\langle \phi _{N}|T|\phi _{N}\rangle }{\langle \phi
_{N}|W|\phi _{N}\rangle },
\]
with $T$ the total kinetic energy operator and $W=\Sigma
_{i=1}^{3}r_{i}\partial V/\partial r_{i}.$ It is known\cite{Brasden} that
for a system of particles interacting through the Coulomb interaction this
quantity has to be 1 for the exact wave function. We obtained a value of
0.999 for almost all the quantum well widths studied, which suggests that
our wave function is well chosen.

\begin{figure}[tbp]
\begin{center}
\epsfig{file=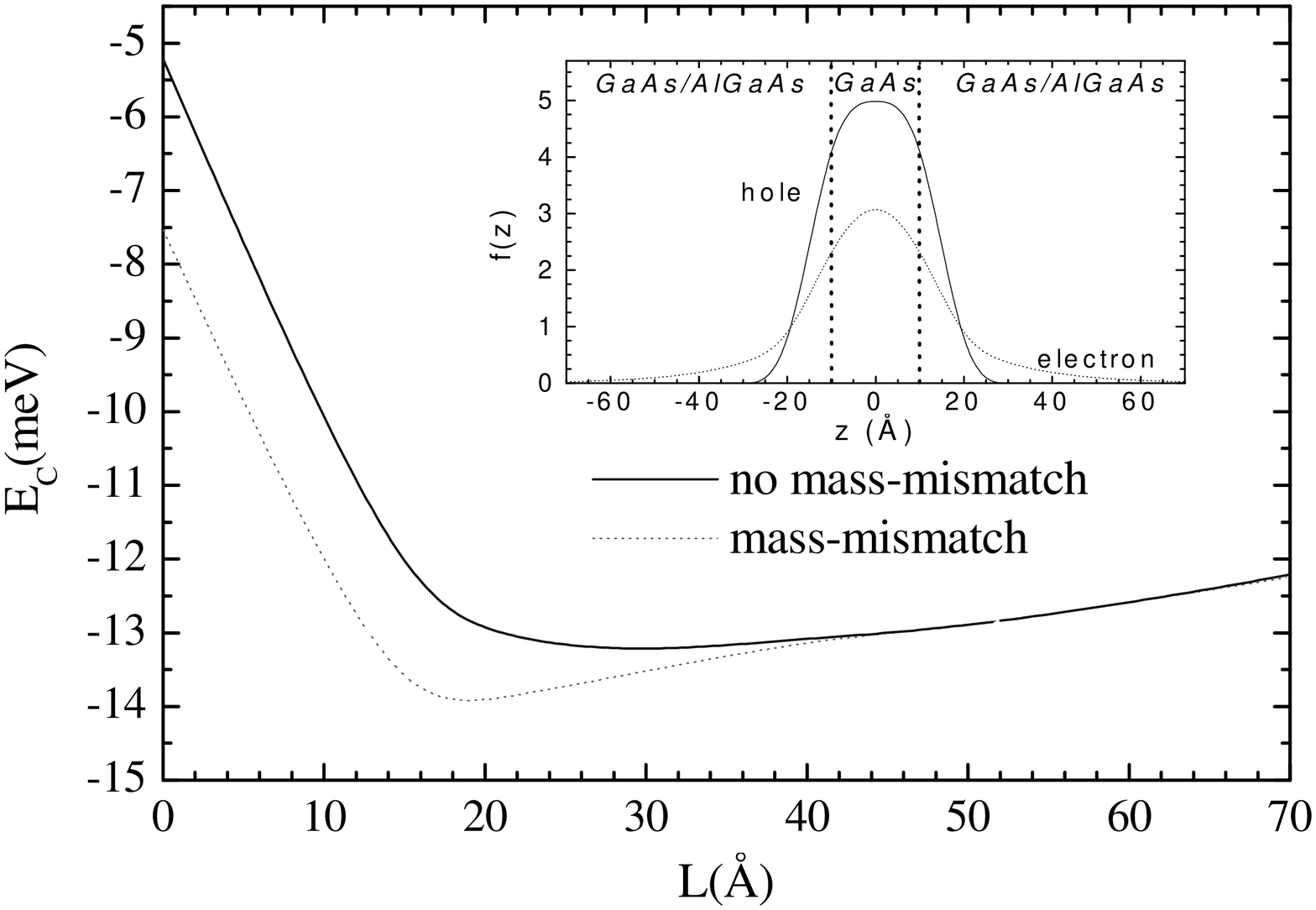,width=10cm}
\end{center}
\caption{The correlation energy of the negative charged exciton vs. the
quantum well width, for the case of constant masses in the well and in the
barrier, and for the the case of different electron and hole masses in the
well and in the barrier of a GaAs/Al$_{0.3}$Ga$_{0.7}$As quantum well. In
the inset the wave function for both the electron and the hole are shown.}
\label{asymm}
\end{figure}

Next we investigate the effect of taking a different mass of the particles
in the well (GaAs) and in the barrier (Al$_{x}$Ga$_{1-x}$As) material, which
is expected to be important in the narrow well regime where the electron and
hole wave functions penetrate into the barrier (see inset of Fig. \ref{asymm}%
). The values for the GaAs-masses, i.e. the masses for the electron and the
hole, are taken equal to the one used in the previous calculation. The
values for the masses in Al$_{x}$Ga$_{1-x}$As are $m_{eb}^{\ast
}=0.067+0.083x$, $m_{hb}^{\ast }=0.34+0.42x$, where $x$ indicates the
percentage of Al present in the alloy. If we assume, as a first
approximation, that part of the electron and the hole wave function is in
the quantum well and the rest is in the barrier we may take the total
effective mass of the electron and the hole as given by
\begin{equation}
\frac{1}{m_{i}}=\frac{P_{iw}}{m_{iw}}+\frac{P_{ib}}{m_{ib}},
\end{equation}
where $m_{iw},m_{ib}$ are the masses of the $i$-th particle in the barrier
and in the well, and $P_{iw}$, $P_{ib}$ are the probabilities of finding the
$i$-th particle in the well and in the barrier, respectively. The results of
this calculation are shown in Fig. \ref{asymm} for $x=0.3$. The correlation
energy increases in absolute value and this is consistent with the fact that
the effective masses are now larger. The effect of the mass mismatch is
important only in the narrow quantum well regime, i.e. L$<$ $40$ \AA , where
it leads to a substantial increase of the magnitude of the correlation
energy. In the L=0 limit we obtain now E$_{C}$(X$^{-}$)$=-$7.5meV which
compares to the 3D correlation energy of a trion in Al$_{0.3}$Ga$_{0.7}$As
which we found to be E$_{C}^{3D}$(X$^{-}$)$=-$6.6 meV. The minimum of the
correlation energy is now obtained at L=17 \AA .

\begin{figure}[tbp]
\begin{center}
\epsfig{file=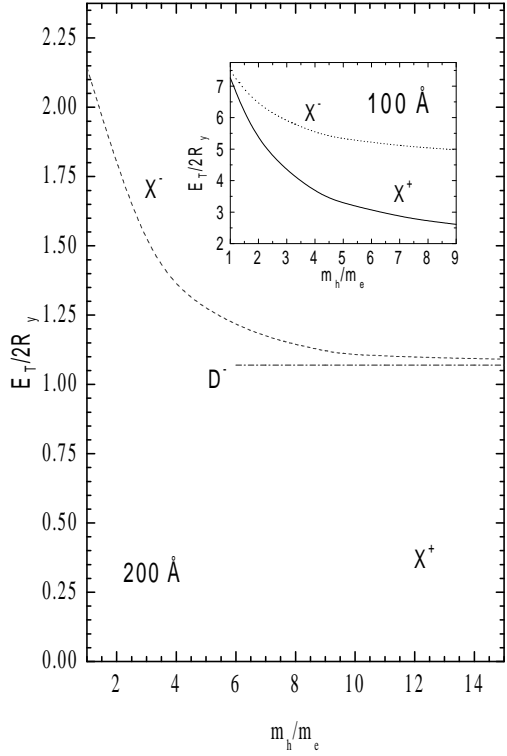,width=8cm}
\end{center}
\caption{The total energy of the negative charged exciton and the positive
charged exciton vs. $m_{h}/m_{e}$ for a 200 \AA\ wide quantum well and for a
quantum well of width 100 \AA\ (inset). The total energy of a D$^{-}$ in the
same quantum well is given by the dash-dotted line for comparison.}
\label{boh}
\end{figure}

We also studied the dependence of the total energy on the hole mass for a
100 \AA\ and a 200 \AA\ wide quantum well. The result, reported in Fig. \ref
{boh}, shows that the total energy decreases as the hole mass increases. The
energy of the negatively charged exciton approaches the energy of the $D^{-}$
in the same quantum well from above and they become practically equal when $%
m_{h}/m_{e}>$ 16 for the 200 \AA\ quantum well. Note that for large values
of the hole mass the X$^{+}$ energy is practically parallel to the one of
the X$^{-}$. In fact, if the hole mass is large its confinement energy
contribution to the total energy is negligible, and the difference between
the X$^{+}$ and X$^{-}$ total energies is just the confinement energy of one
electron, which of course does not depend on the hole mass.

\begin{figure}[tbp]
\begin{center}
\epsfig{file=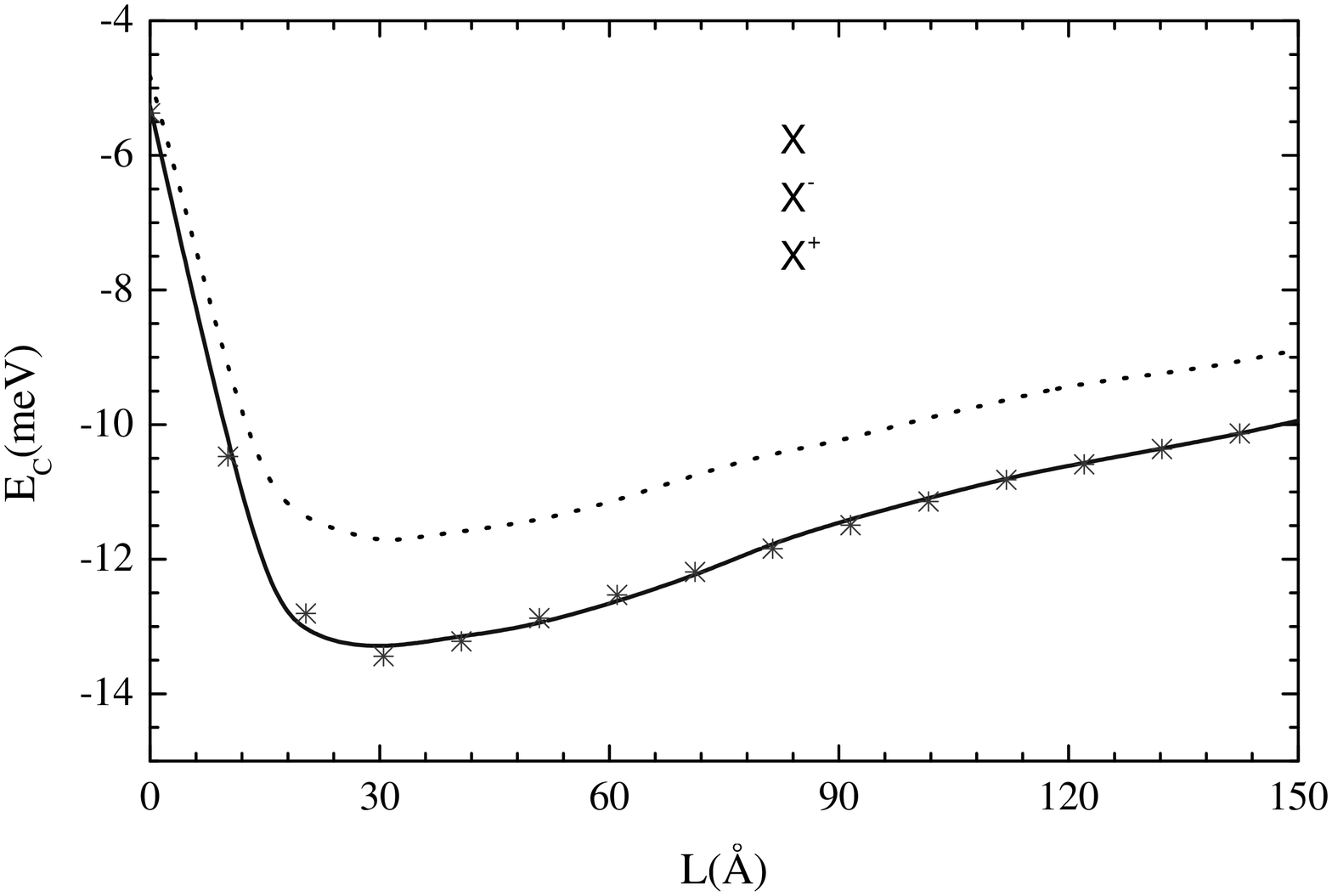,width=8cm}
\end{center}
\caption{The correlation energy of the negatively charged exciton (symbols)
and the one of the positively charged exciton (solid curve) vs. the well
width. The correlation energy of the exciton (dotted curve) is given as
reference.}
\label{XmVsXp}
\end{figure}

Next we studied the correlation energy of the positively charged exciton. In
Fig. \ref{XmVsXp} we plot the correlation energy of the X$^{-}$ and X$^{+}$
systems as function of the well width. Note that the correlation energy of
the X$^{+}$ is equal to the one of X$^{-}$ (within the numerical accuracy).
This is in agreement with recent experimental data\cite{Finkelstein} where
the binding energy of the X$^{+}$ was found to be equal to the one of \ the X%
$^{-}$. In fact we have
\begin{eqnarray}
\text{E}_{C}\text{(X}^{-}\text{) }-\text{ E}_{C}\text{(X}^{+}\text{)} &=&%
\text{E(X}^{-}\text{) }-\text{ E}_{e}+\text{ E}_{h}-\text{ E(X}^{+}\text{)}
\nonumber \\
\text{{}} &=&\text{[E(X}^{-}\text{) }-\text{E(X)}-\text{E}_{e}\text{] }-%
\text{ [E(X}^{+}\text{) }-\text{ E(X) }-\text{E}_{h}\text{]}  \nonumber \\
\text{{}} &=&\text{E}_{B}\text{(X}^{-}\text{) }-\text{ E}_{B}\text{(X}^{+}%
\text{),}
\end{eqnarray}
where E$_{B}$(X$^{\pm }$) is the binding energy of a charged exciton system
referred to the one of the exciton plus one free electron (hole) system,
\begin{equation}
\text{E}_{B}\text{(X}^{\pm }\text{)=E(X) + E}_{h(e)}\text{ - E(X}^{\pm }%
\text{),}  \label{bin}
\end{equation}
where E(X) is the energy of the exciton, E$_{e(h)}$ is the energy of the
free electron (hole) and E(X$^{\pm }$) is the charged exciton binding
energy. Consequently, if the X$^{-}$ and the X$^{+}$ correlation energy are
the same, the corresponding binding energies will also be the same.

\begin{figure}[tbp]
\begin{center}
\epsfig{file=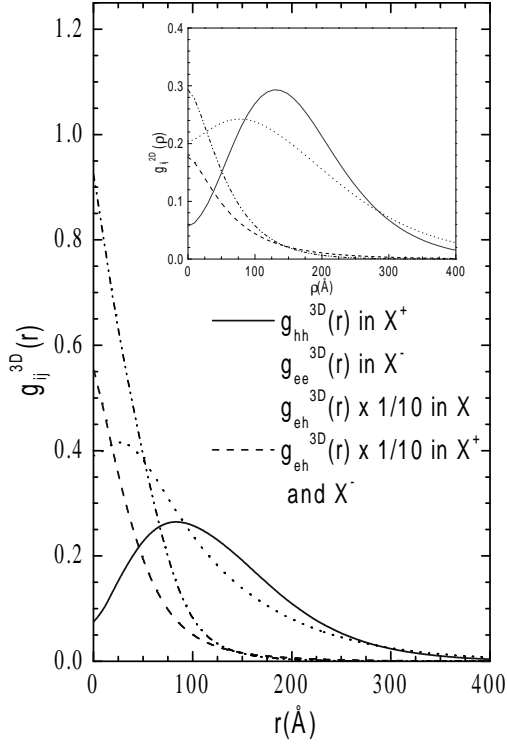,width=8cm}
\end{center}
\caption{3D pair correlation function for different pairs of particles in X$%
^{-}$ and in X$^{+}$ vs. the distance between the particles. In the inset
the 2D pair correlation function is shown. The curve convention is the same
in the two plots. }
\label{pair}
\end{figure}

Last we study the wave function of the negatively and positively charged
exciton and the correlation between the different particles. The pair
correlation function, $g_{ij}^{3D}(r)=\langle \delta \left( r-|%
\overrightarrow{r}_{i}-\overrightarrow{r}_{j}|\right) \rangle $, for a 100
\AA\ wide quantum well is shown in Fig. \ref{pair}. This function gives the
probability to find particle $i$ and particle $j$ at a distance $r$ from
each other. Notice that $g_{eh}^{3D}(r)$ is the same for both X$^{-}$ and X$%
^{+}$ (dashed curve in Fig. \ref{pair}) and in both cases the electron and
the hole tend to be close to each other. A similar result is obtained for
the exciton (dot-dashed curve in Fig. \ref{pair}). The fact that the
intensity of the correlation function for the exciton is higher than the one
of the charged exciton is a direct consequence of the normalization of the
wavefunction to one. The situation is very different for the correlation
between particles having the same charge. For the X$^{-}$ electrons $%
g_{ee}^{3D}(r)$ (dotted curve in Fig. \ref{pair}) shows that the two
electrons avoid each other at small distances and have the highest
probability of sitting at a distance of 25 \AA $\approx a_{B}/4$. For the
holes in X$^{+}$ $g_{hh}^{3D}(r)$ (solid curve in Fig. \ref{pair}) shows
that the two holes avoid each other at small distances and have the highest
probability of sitting at a distance of 80 \AA $\approx 4a_{B}/5$, thus
farther from each other than the electron-couple in X$^{-}$. However the
average distance,$\langle |\overrightarrow{r}_{i}-\overrightarrow{r}%
_{j}|\rangle $, of the two electrons in X$^{-}$, does not differ much from
the average distance between the holes in X$^{+}$. We found 250 \AA\ and 216
\AA\ respectively. The average distance between the electron and the hole is
150 \AA\ and is found to be the same in the X$^{-}$ and in the X$^{+}$. In
the inset of Fig. \ref{pair} we show the 2D-pair correlation function, $%
g_{ij}^{2D}(\rho )=\langle \delta \left( \rho -|\overrightarrow{\rho }_{i}-%
\overrightarrow{\rho }_{j}|\right) \rangle $, for a 100 \AA\ wide quantum
well, where the same curve conventions are used as for the 3D pair
correlation functions. These 2D correlation functions express more clearly
the Coulomb correlation between the particles. In the 3D correlation
functions the z-direction is still involved . In this direction the quantum
well potential forces the particles towards the middle of the well. As a
consequence all 2D correlation functions are more spread out as compared to
their 3D counterpart. The peaks in the electron-electron and hole-hole
correlation functions are shifted towards larger distances. The average
distances in the $\rho $-plane of the electrons in X$^{-}$ and of the holes
in X$^{+}$ is $\left\langle \left| \overrightarrow{\rho }_{i}-%
\overrightarrow{\rho }_{l}\right| \right\rangle =$ 249 \AA\ and 214 \AA ,
respectively. This result differs only by a few angstroms as compared to the
3D result, suggesting that the charged exciton, for L=100 \AA , is almost
bidimensional.

\begin{figure}[tbp]
\begin{center}
\epsfig{file=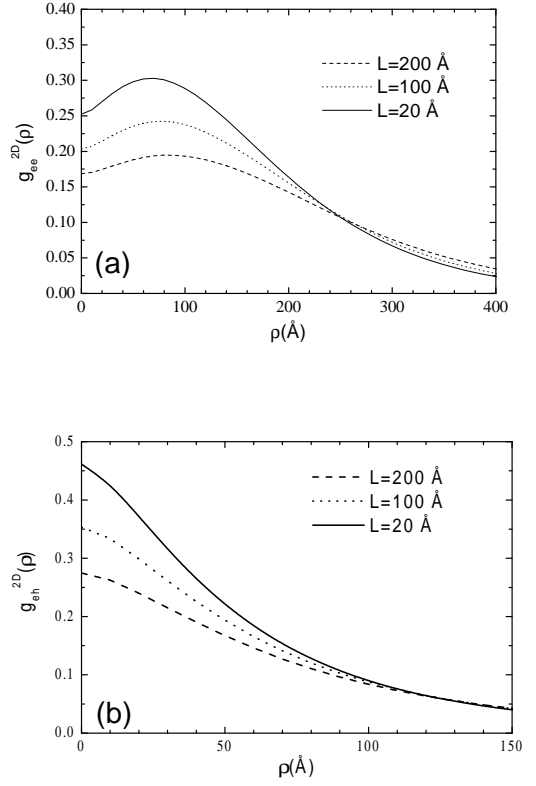,width=8cm}
\end{center}
\caption{The pair correlation function for the electron-electron (a) and
electron-hole (b) in X$^{-}$ for different quantum well widths . }
\label{function-L}
\end{figure}

We now look at the 2D correlation function for different  well widths (see
Fig. \ref{function-L}). Notice that the peak of the correlation function for
the electron-electron couple in X$^{-}$ slowly shifts towards smaller
distances as the well width decreases (see Fig. \ref{function-L}(a)), at the
same time the tail of the function becomes smaller. The peak of the
electron-hole correlation function also increases (see Fig. \ref{function-L}%
(b)) but is still centered around zero. With decreasing well width the X$^{-}
$ becomes less extended. A similar behavior was observed for X$^{+}.$

\begin{figure}[tbp]
\begin{center}
\hspace{-2cm}\epsfig{file=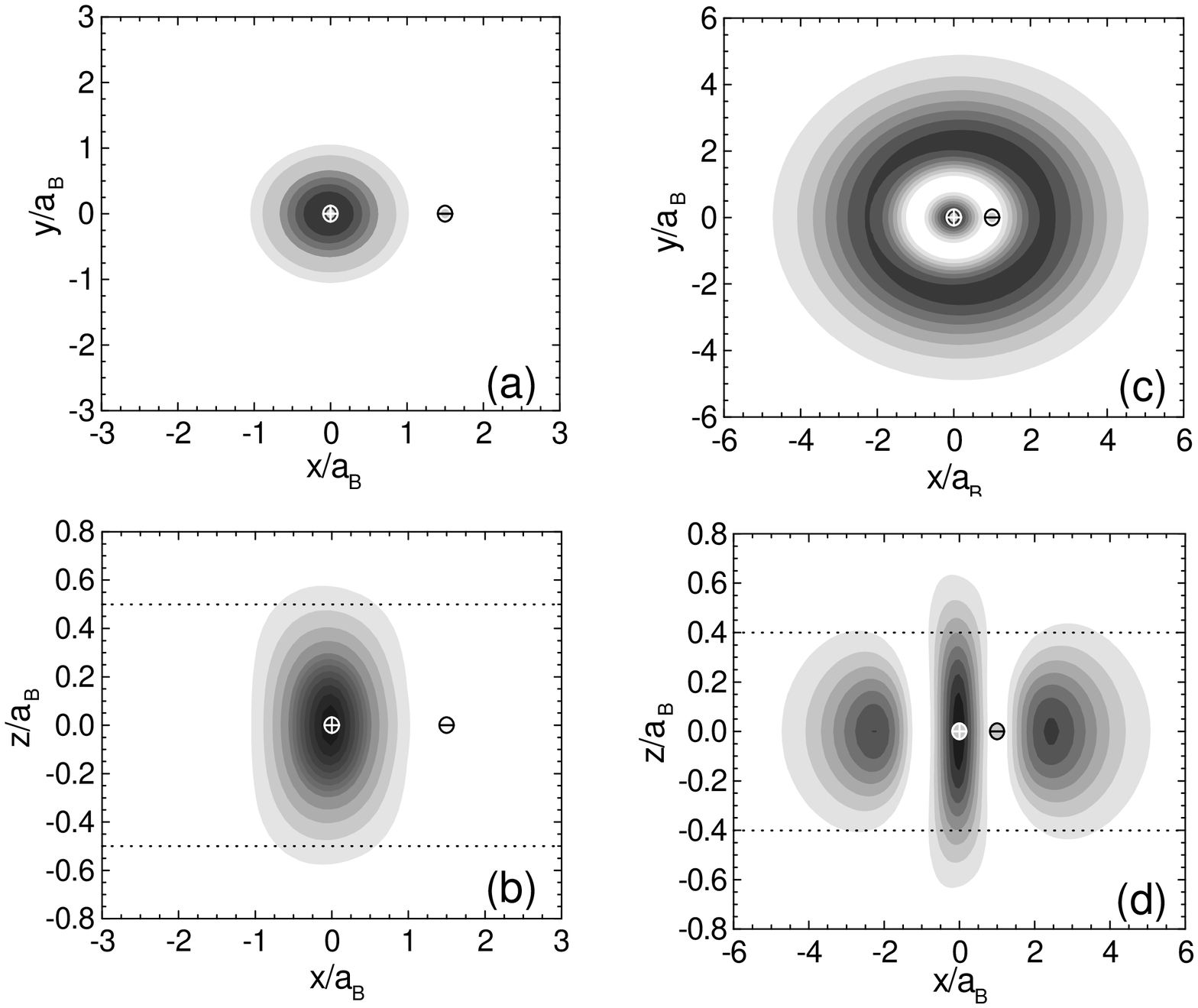,width=10cm}
\end{center}
\caption{Contour maps of the conditional probability for the X$^{-}$ in a
quantum well of width 100\AA $\approx $ a$_{B}$. The fixed particles are
indicated by symbols (circle with a cross for the hole and circle with a
minus sign for the electron). The dotted lines indicate the quantum well
boundaries. }
\label{geometry}
\end{figure}

In Fig. \ref{geometry} we show the contour plots of $|\phi _{0}(%
\overrightarrow{r}_{1e},\overrightarrow{r}_{2e},\overrightarrow{r}_{h})|^{2}$
for a negatively charged exciton in a quantum well of width 100 \AA , where
lighter regions correspond to lower probability. In Figs. \ref{geometry}%
(a,b) we plot the projection of the electron probability density in the xy-
and xz-plane when the hole is fixed at $\overrightarrow{r}_{h}=(0,0,0)$ and
one of the two electrons is fixed at $\overrightarrow{r}_{1e}=(1.5a_{B},0,0)$%
. The distance between the particles is equal to the average electron-hole
distance in the X$^{-}.$ The two symbols show the positions of the two fixed
particles. The second electron sits close to the hole and the fixed electron
sees an exciton consisting of the hole-second electron couple. Notice, from
Fig. \ref{geometry}(b), that the second electron slightly penetrates the
barrier. In Figs. \ref{geometry}(c,d) we plot the projection of the
probability density in the xy- and xz-plane when the hole is fixed at $%
\overrightarrow{r}_{h}=(0,0,0)$ and one of the two electrons is fixed at $%
\overrightarrow{r}_{1e}=(a_{B},0,0)$. Thus, the distance between the
electron and the hole is now smaller than the average distance. Now, in the
xy-plane{\em \ }a small part of{\em \ }the second electron sits on top of
the hole and the largest part is situated outside the white ellipse defined
by the position of the fixed electron. Thus now the fixed electron and hole
act like an exciton to which the second electron is bound. The situation is
similar in the xz-plane where part of the second electron sits on top of the
hole and the other part is almost symmetrically distributed in two puddles
around x=$\pm 2.2a_{B}.$

\begin{figure}[tbp]
\begin{center}
\epsfig{file=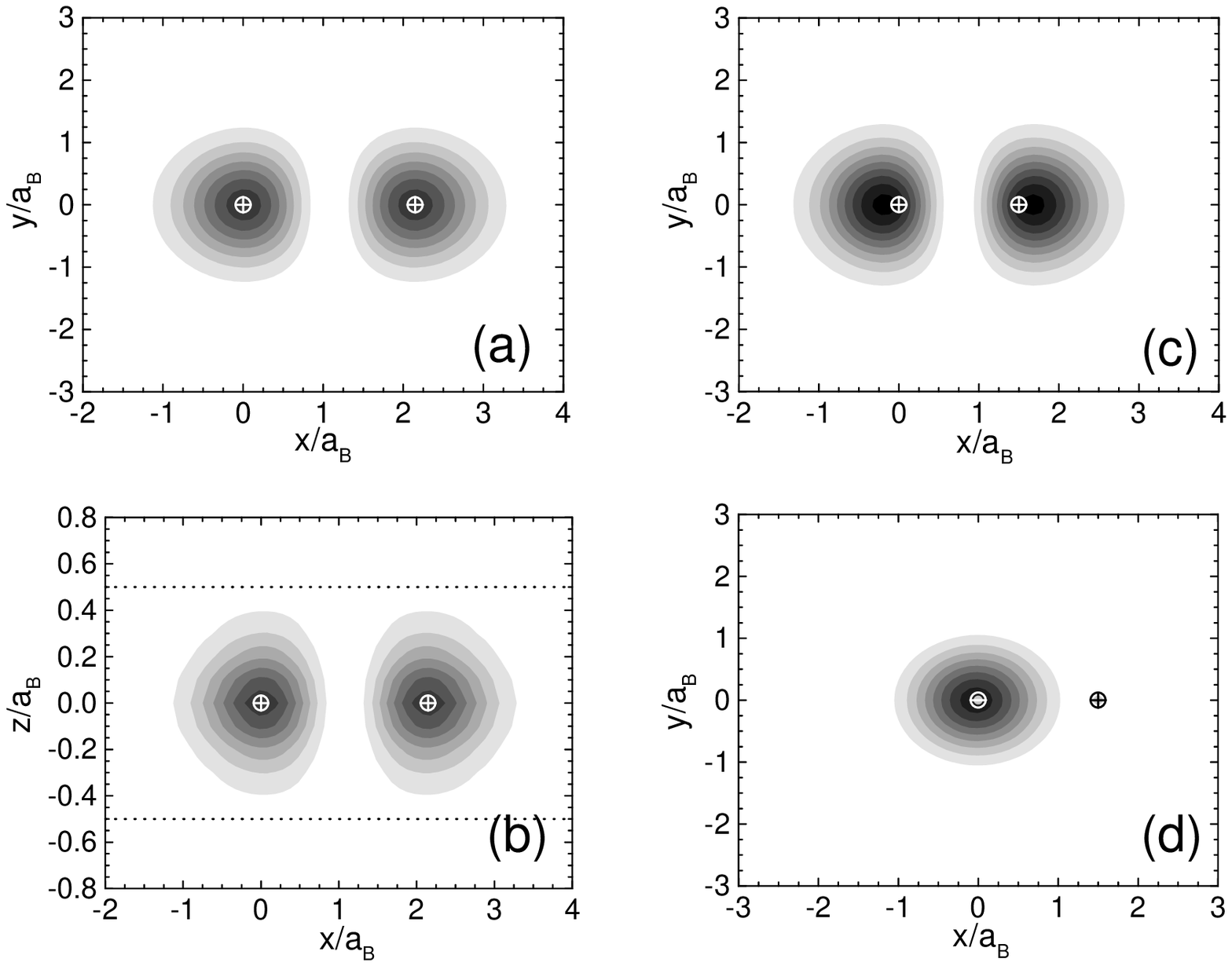,width=12cm}
\end{center}
\caption{Contour maps of the conditional probability for the X$^{+}$ in a
quantum well of width 100 \AA $\approx $a$_{B}$. The fixed particles are
indicated by symbols (circle with a cross for the hole and circle with a
minus for the electron). The dotted lines indicate the quantum well
boundaries. }
\label{geometry2}
\end{figure}

In Fig. \ref{geometry2} we show the contour plots of $|\phi _{0}(%
\overrightarrow{r}_{1h},\overrightarrow{r}_{2h},\overrightarrow{r}_{e})|^{2}$
for a positively charged exciton in a quantum well of width 100 \AA . In
Figs. \ref{geometry2}(a,b), we plot the projection of the electron
probability density function on the xy- and the xz-plane where we fixed the
two holes at $\overrightarrow{r}_{1h}=(0,0,0)$ and $\overrightarrow{r}%
_{2h}=(2.2a_{B},0,0).$ The distance between the holes is now equal to their
average distance. Notice that the electron is now equally distributed over
the two holes. Notice also (see Fig. \ref{geometry2}(b)) that the electrons
do not penetrate into the barrier, as opposite to what happens for the X$%
^{-} $, indicating that the electron is now more strongly bound. In Fig. \ref
{geometry2}(c), we show again the projection of the probability density
function on the xy{\em -}plane, where now the second hole is fixed at $%
\overrightarrow{r}_{2h}=(1.5a_{B},0,0).$ In this case the lobes of the
electron probability density function repel each other and are no longer
centered on the position of the holes. The holes are closer than their
average distance but the distance among the centers of the lobes is still
approximately equal to the hole-hole average distance in the X$^{+}$. In
Fig. \ref{geometry2}(d) we fix the position of the electron $\overrightarrow{%
r}_{e}=(0,0,0)$ and the first hole $\overrightarrow{r}_{1h}=(1.5a_{B},0,0),$
such that the average distance is equal to the electron-hole pair average
distance. The second hole is now centered around the electron which is the
same as the behavior of the second electron in X$^{-}$(see Fig. \ref
{geometry}(a)). We found that the contour plot of the probability density of
finding the hole in a X$^{-}$ at a position $\overrightarrow{r}$ when the
two electrons are fixed, is practically the same as the one for the electron
in the X$^{+}$ with the two holes fixed (see Fig. \ref{geometry2}(a)).

\begin{figure}[tbp]
\begin{center}
\epsfig{file=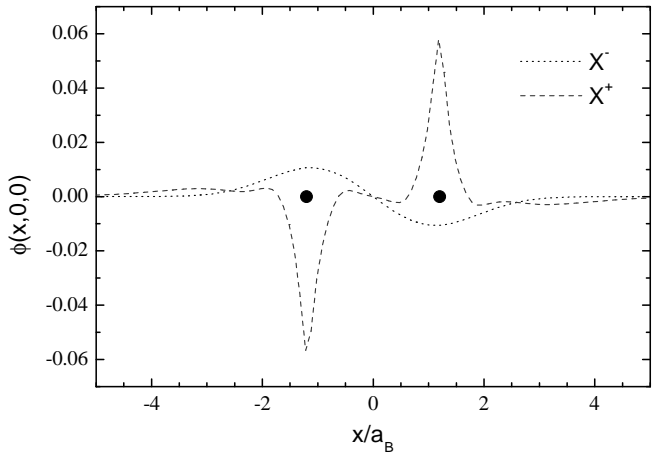,width=10cm}
\end{center}
\caption{The wavefunction of the hole in X$^{-}$ when the two electrons are
fixed (dotted curve) and of the electron in X$^{+}$ when the two holes are
fixed (dashed curve) along the direction [1,0,0]. The two solid dots
indicate the two equally charged fixed particles. }
\label{fig-wave}
\end{figure}

In Fig. \ref{fig-wave} we show the wavefunction of the hole in X$^{-}$ and
the electron in X$^{+}$ along the $x$-axis when the two particles having the
same charge are fixed (solid circles in Fig. \ref{fig-wave}). Notice that
although the trions are in a singlet state the wavefunction is
anti-symmetric for reflections around the mid-point between the two fixed
particles. The reason is that  interchanging the two fixed particles must
result in a sign change of the wavefunction. Remark that the electron in X$%
^{+}$ is more localized on the two holes as compared to the hole in X$^{-}$
which is spread out over the two electrons. However in both cases the
wavefunction has a node between the two fixed particles, in contrast to what
would happen if the interaction among the particles would be ``chemical
bonding''-like. It seems then reasonable to say that in the same way in
which the X$^{-}$ can be described as an exciton with an extra electron
moving around the electron-hole couple and weakly bound to it, the X$^{+}$
can be viewed as an exciton in which an extra hole moves in a orbit around
the electron-hole couple. The latter picture is different from a system in
which two holes bind through an electron, i.e. H$_{2}^{+}$-like. Another
confirmation for this picture comes from the pair correlation functions.
Suppose that the electron, for X$^{-}$, is in the origin, then the hole will
be near the origin as indicated by the electron-hole correlation function
and the other electron will be situated around the position of the peak in
the electron-electron pair correlation function. So the picture we get is
an electron-hole pair with an extra electron moving around
it. If we switch the role of hole and electron, a similar picture can be
imagined for the X$^{+},$ with the only difference being that now the extra
hole sits even further  from the electron-hole couple than in the X$^{-}$.
Thus, the charged exciton is similar to the charged positron. The similarity
in the structure of the two different species of charged excitons is
consistent with the fact that their correlation energy is found to be equal.

\section{Comparison with experiments}

\begin{figure}[tbp]
\begin{center}
\epsfig{file=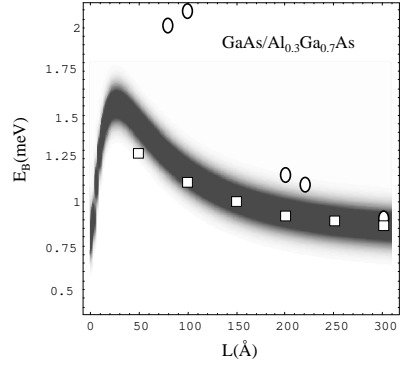,width=8cm}
\end{center}
\caption{The theoretical (shaded curve) and experimental (open circles)
binding energies of the negatively charged exciton in a GaAs/AlGaAs quantum
well vs. the well width. The theoretical results of the quantum Monte Carlo
calculation of Ref.22 are shown by the open squares.}
\label{binding}
\end{figure}

Experimental data in zero magnetic field were reported for the binding
energy of the X$^{-}$ in a 100 \AA \cite{Kaur}, 200 \AA \cite{Finkelstein}\
, 220 \AA \cite{Shields95}\ and in a 300 \AA \cite{Shields95}\ quantum well.
The reported values are 2.1, 1.15,\ 1.1$\pm 0.1$ and 0.9$\pm 0.1$
meV\thinspace\ respectively and are compared in Fig. \ref{binding} with our
theoretical results. The value of E$_{B}=2$ meV\ for a 80 \AA\ well is for a
GaAs/AlAs quantum well and was measured by Yan {\em et al.}\cite{Govaerts}.
The theoretical results for the binding energy are represented by a shaded
region which gives the accuracy of our calculation for the binding energy.
Note that the accuracy obtained for the total energy is better than 1\%,
however, its error propagates and increases because of the subtractions (see
Eq. (\ref{bin})) which have to be made in order to obtain the binding
energy, which is one order of magnitude lower than the total energy. An
important consequence of this observation is that any approximation made in
the calculation of $E(X^{\pm })$ may lead to substantial errors in the
binding energy. For comparison we also report (open squares) the theoretical
result obtained by Tsuchiya and Katayama\cite{Japan} using the quantum Monte
Carlo method. Notice that the results of Ref.\onlinecite{Japan} agree very
well with ours, however our calculation goes down to smaller well width. The
binding energy first increases with increasing quantum well width and then,
after reaching a maximum of $E_{B}$=1.6meV at L $\thickapprox $ 35 \AA\ %
starts to decrease. The decrease becomes very slow for quantum well width
above 70\AA . The increase of the binding energy with decreasing quantum
well width agrees qualitatively with the experimental data, but the
experimental increase is much faster than the one we find theoretically. The
inclusion of the conduction band non-parabolicity would increase the binding
energy only slightly. We believe that the increased discrepancy between
theory and experiment with decreasing well width is a consequence of the
localization of the trion due to the presence of quantum well width
fluctuations, as was also found for biexcitons\cite{clara-bi,Oliver}. This
is consistent with the fact that for L=300\AA\ our result agrees with the
experiments and that the effect of the quantum well width modulation on the
localization of the exciton and the trion increases with decreasing well
width.

\begin{figure}[tbp]
\begin{center}
\epsfig{file=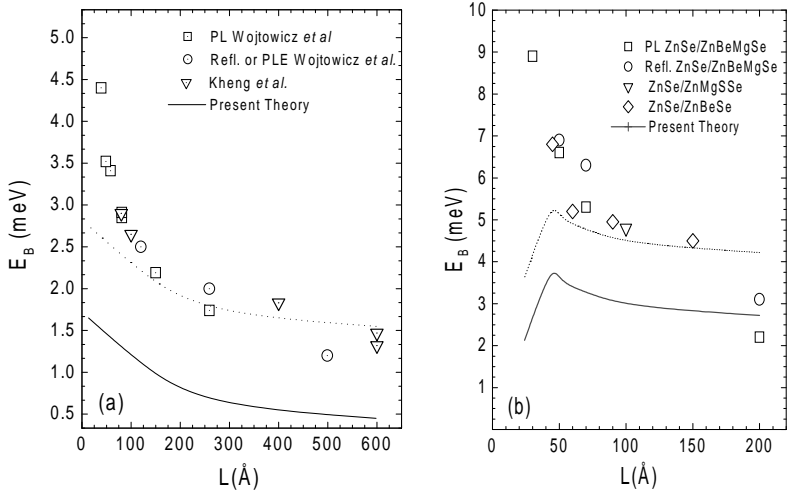,width=11cm}
\end{center}
\caption{The binding energy of the negative charged exciton vs. the quantum
well width for the charged exciton for CdTe-based structures (a) and for
ZnSe-based structures (b). The experimental data are taken from Refs.
2,24, for the CdTe-based structures, and from Ref.
25 for the ZnSe-based structures. Our theoretical results are
given by the solid curve. The dotted curve is our result shifted by a
constant.}
\label{CdTe}
\end{figure}

A similar calculation was done for CdTe quantum well structures (Fig. \ref
{CdTe}(a)) and ZnSe based quantum well structures (Fig. \ref{CdTe}(b)). The
binding energy versus the well width in these materials is shown in Figs.
\ref{CdTe}(a,b) (solid curve) and is compared to experimental data \cite
{Kheng,cdte,znse} (symbols in Figs. \ref{CdTe}(a,b)). The parameters used in
the calculation for CdTe based structures are V$_{e}=216.4$ meV, V$_{h}=163$
meV, m$_{e}$=0.096m$_{0}$, m$_{h}$=0.19m$_{0}$, a$_{B}=54$ \AA , which
results into R$_{y}=13.8$ meV. The value of the barriers are taken from Ref.%
\onlinecite{cdte}. Notice that for this structure we have the same potential
barrier heights than for the GaAs case, however the ratio between the mass
of the electron and the mass of the hole is very different, namely m$_{e}/$m$%
_{h}=0.505$ (CdTe) as compared to m$_{e}/$m$_{h}=0.196$ (GaAs). In the range
200\AA\
\mbox{$<$}%
L
\mbox{$<$}%
600 \AA\ the theoretical curve is shifted by about 1 meV with respect to the
experimental results. Below 200\AA\ the experimental results increase faster
with decreasing L\ as compared to the experimental data which is probably a
consequence of the above mentioned increased localization of the trion.

For the ZnSe structure we use the parameters of the ZnSe/ZnBeMgSe
structures, $\Delta $V$=230$ meV with V$_{e}=0.70\Delta $V, V$%
_{h}=0.30\Delta $V, m$_{e}=0.16$m$_{0},$ m$_{h}=0.8$m$_{0},$ a$_{B}=30.05$
\AA , which results in R$_{y}=53.34$ meV. The results are qualitatively
similar to the one obtained for the GaAs/AlGaAs quantum wells. The agreement
with experiment is also in this case not satisfactory (see Fig.\ref{CdTe}%
(b)) except for the 200 \AA\ wide quantum well.

To understand the fact that the theoretical results for CdTe- and ZnSe-based
structures underestimate so much the experimental data even for large well
widths, we have to take into account that these materials are strongly
polar. In the present work we are neglecting polaronic effects and it is
known, at least for the case of excitons, that this leads to an
underestimation of the binding energy of the system\cite{nardi}. Currently
only a calculation of the polaron correction to the ground state of a D$^{-}$
system is available\cite{Shi} but no calculation for the trion system has
been published. For the D$^{-}$system we know that the polaron correction
equals the 3D polaron correction down to rather small well widths. When, we
shift the results in Figs. \ref{CdTe}(a,b) by the constant values 1.1meV and
1.6 meV, respectively (dotted curve in Figs. \ref{CdTe}(a,b), they agree
very well with the experimental results over a large range of quantum well
widths. We believe that these shifts are due to polaron effects. Shi {\em et
al.}\cite{Shi} obtained an upper limit of $\approx 0.4\alpha \hbar \omega
_{LO}$ to the polaron contribution to the binding energy of the D$^{-}$
system in a wide quantum well ($\alpha $ is the electron-phonon coupling
constant and $\hbar \omega _{LO}$ is the optical phonon energy). In an X$^{-}
$ system the hole is not localized which will strongly reduce the polaron
effect to an estimated value of 0.1-0.2 $\alpha \hbar \omega _{LO}.$ For
CdTe quantum wells with $\alpha =0.3$ and $\hbar \omega _{LO}=21.1$ meV this
gives 0.6-1.2 meV while for ZnSe\cite{Shi2} we have $\alpha =0.42$ and $%
\hbar \omega _{LO}=31.5$ meV and consequently 1.3-2.6 meV. These value are
comparable to the shifts in Fig. \ref{CdTe}(a,b), and agree with the fact
that the shifts for the ZnSe quantum wells is larger than for CdTe quantum
wells.

\section{Conclusion}

In this paper we applied the stochastic variational method to study the
ground state of the exciton and the charged exciton in a quantum well. This
is the first time, to our knowledge, that a calculation fully includes the
effect of the Coulomb interaction and the confinement due to the quantum
well, and thus the particle-particle correlation in both the direction of
the quantum well and the confinement direction. The results obtained do not
show a big qualitative difference from the one already present in the
literature, however substantial quantitative differences are found. This
difference leads to an improvement in the agreement with experimental data.
However, the experimentally measured binding energy for a negatively charged
exciton increases much faster with increasing well width at small well width
than our theoretical results. We believe that this discrepancy is a
consequence of the increased localization of the exciton and trion with
decreasing well width. A similar conclusion was also reached recently for
biexcitons \cite{Oliver,clara-bi}. For CdTe- and ZnSe-based quantum wells
the polaron effect, which was not included in our approach, is expected to
lead to a substantial shift ($\sim 1$ meV) of the binding energy to larger
values. Also in this case the trapping of the trions on the quantum well
width fluctuations is probably responsible for the rapid increase of the
trion binding energy below L$\approx 100$ \AA\ . The study of the
conditional probability distribution of the particles in the system and of
the pair correlation functions lead us to conclude that a charged exciton is
similar to a charged positron. This conclusion is important as it supports
the fact that the correlation energy for X$^{-}$ and X$^{+}$ is found to be
equal.

\section{Acknowledgment}

Part of this work is supported by the Flemish Science Foundation (FWO-Vl)
and the `Interuniversity Poles of Attraction Program - Belgian State, Prime
Minister's Office - Federal Office for Scientific, Technical and Cultural
Affairs'. F.M.P. is a Research Director with FWO-Vl. Discussions with M.
Hayne and correspondence with B. St\'{e}b\'{e} are gratefully acknowledged.
K. Varga was supported by the U. S. Department of Energy, Nuclear
Physics Division, under contract No. W-31-109-ENG-39 and OTKA grant
No. T029003 (Hungary).

\pagebreak

\pagebreak

\end{document}